\documentclass[
reprint,
 amsmath,amssymb,
 aps,
]{revtex4-1}

\usepackage{graphicx}
\usepackage{dcolumn}
\usepackage{bm}
\usepackage{array}
\usepackage{hyperref}
\hypersetup{colorlinks,%
citecolor=blue,%
urlcolor=blue,%
pdftex}
\usepackage{subfigure}

\begin{document}

\preprint{APS/123-QED}

\title{Berry curvature in monolayer MoS$_{2}$ with broken mirror symmetry}%

\author{Kyung-Han Kim}
\affiliation{Department of Physics, Pohang University of Science and Technology, Pohang 37673, Korea}
\author{Hyun-Woo Lee}%
\affiliation{Department of Physics, Pohang University of Science and Technology, Pohang 37673, Korea}




\date{\today}

\begin{abstract}

An ideal 1H phase monolayer MoS$_{2}$ has the mirror reflection symmetry but this symmetry is broken in common experimental situations, where the monolayer is placed on a substrate.
By using the k$\cdot$p perturbation theory, we investigate the effect of the mirror symmetry breaking on the Berry curvature of the material.
We find that the symmetry breaking may modify the Berry curvature considerably and the spin/valley Hall effect due to the modified Berry curvature is in qualitative agreement with a recent experimental result [Science {\bf 344}, 1489 (2014)], which cannot be explained by previous theories that ignore the mirror symmetry breaking.

\end{abstract}

\maketitle



\section{Introduction}\label{sec:introduction}

In solids of two-dimensional (2D) hexagonal structure, an electron has not only spin but also valley degree of freedom, which acts as a pseudospin.
The spin can be used for information storage, transport, and manipulation, and is the central degree of freedom for spintronics~\cite{S. A. Wolf 2001,I. Zutic 2004}.
It was recently realized~\cite{Q. H. Wang 2012,X. Xu 2014,R. Ganatra 2014,O. Bleu 2017} that the valley can play similar roles as the spin, opening the field of valleytronics.
How to control the spin/valley degree of freedom is one of fundamental questions in spin/valleytronics, and the spin/valley Hall effect (SHE/VHE) is one possible way to achieve such control.

A prototypical material of 2D hexagonal structure is graphene, which has been studied extensively.
Recently monolayer transition metal dichalcogenides (TMD) also have attracted huge attention as a 2D hexagonal meterial. Unlike the graphene, a monolayer TMD may have direct bandgap with suitable gap size and large spin-orbit coupling (SOC). It is thus a good candidate material for optoelectronic and spin/valleytronic devices~\cite{Q. H. Wang 2012}.

\begin{figure}[ht] 
	\centering 
	\includegraphics[angle=0, width=8.5cm, height=4cm]{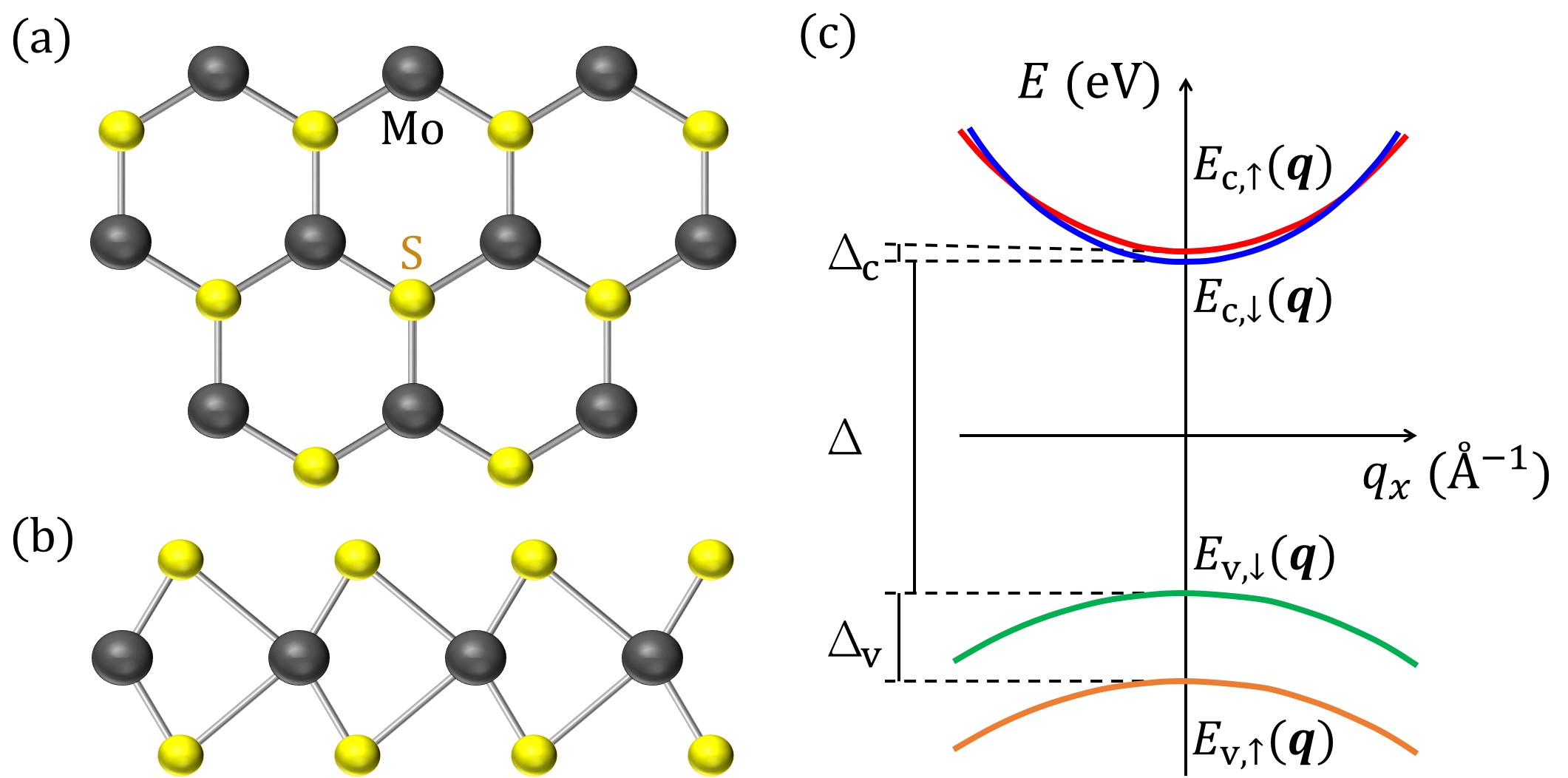} 
	\subfigure{\label{fig:1a}}
	\subfigure{\label{fig:1b}}
	\subfigure{\label{fig:1c}}
	\caption{\label{fig:1}
		(Color online) (a) Top and (b) side view of an ideal 1H phase monolayer MoS$_{2}$, which consists of Mo atoms (larger dot) and S atoms (smaller dot) and has the mirror symmetry with respect to the middle sublayer plane.
		(c) Schematic band structure near the K point of an ideal monolayer MoS$_{2}$ with the mirror reflection symmetry. Only the two lowest conduction bands and the two highest valence bands are shown.
		Band gap energy is $\Delta=E_{\text{c},\downarrow}(0)-E_{\text{v},\downarrow}(0)\approx1.82\,\text{eV}$ and spin splitting energies are $\Delta_{\text{c}}\approx3\,\text{meV}$ and $\Delta_{\text{v}}\approx148\,\text{meV}$ for conduction and valence bands, respectively. About $70\,\text{meV}$ above $E_{\text{c},\downarrow}(0)$, $E_{\text{c},\uparrow}(\bm{q})$ and $E_{\text{c},\downarrow}(\bm{q})$ become degenerate.} 
\end{figure}

A 1H phase monolayer MoS$_{2}$ [Figs.~\ref{fig:1a},\ref{fig:1b}] is probably most popular among the monolayer TMD materials. In this material, the mirror reflection symmetry is respected but the inversion symmetry is broken intrinsically. Energy bands are split by the SOC with electron spins quantized along the out-of-plane direction. Recent experiments on the monolayer MoS$_{2}$ investigated SHE/VHE~\cite{K. F. Mak 2014,C. Cheng 2016}, spin-orbit torque~\cite{W. Zhang 2016,Q. Shao 2016}, valley magnetoelectric effect~\cite{J. Lee 2017}, valley relaxation~\cite{H. Z. Lu 2013,Q. Wang 2013,C. Mai 2014,S. D. Conte 2015}, and spin relaxation~\cite{L. Wang 2014, L. Yang 2015, Y. J. Zhang 2017}.
Unfortunately many experimental results remain unexplained, which motivates further theoretical studies on this material.

In this paper, we investigate the SHE and VHE in a monolayer MoS$_{2}$ with the broken mirror symmetry. While the mirror reflection with respect to the mirror plane [Fig.~\ref{fig:1b}] within the monolayer MoS$_{2}$ may be a good symmetry for an ideal MoS$_{2}$ monolayer suspended in air, this symmetry is broken in common experimental situations where a MoS$_{2}$ monolayer is placed on a substrate and subject to a gate voltage.
We demonstrate that the mirror symmetry breaking may significantly modify the monolayer’s Berry curvature, which is an important source of the SHE/VHE~\cite{D. Xiao 2010}. This provides an explanation as to why the experimental results~\cite{K. F. Mak 2014,J. Lee 2017} on the SHE/VHE deviate from predictions of the previous theoretical studies~\cite{D. Xiao 2012} that assume the mirror symmetry.
We calculate the spin and valley Hall conductivities as a function of the mirror symmetry breaking strength, which may be continuously modulated in experiments by applying a gate voltage. Qualitative agreement with recent experiments~\cite{K. F. Mak 2014,J. Lee 2017} is found.

The paper is organized as follows.
In Sec.~\ref{sec:theorya}, we introduce the k$\cdot$p perturbed Hamiltonian near the K point of the monolayer MoS$_{2}$ with the mirror symmetry.
In Sec.~\ref{sec:theoryb}, we consider the mirror symmetry breaking effect in terms of the effective Hamiltonian.
In Sec.~\ref{sec:result}, we calculate the Berry curvature and orbital magnetic moment using the effective Hamiltonian and compare our result to recent experimental results.
In Sec.~\ref{sec:discussion}, we discuss various technical issues related with this paper.
Finally, our main results are summarized in Sec.~\ref{sec:summary}.




\section{Theory}\label{sec:theory}

A monolayer MoS$_{2}$ has a direct band gap at the K and K$'$ points~\cite{K. F. Mak 2010}. Since these two points are the time reversed images of each other, study on one point, say the K point, is sufficient to understand properties of the both points. We thus study only the K point which is highly symmetric and has the property of $C_{3\rm h}$ point group.
From the irreducible representations of $C_{3\rm h}$ point group~\cite{A. Kormanyos 2013}, K point basis functions of the two lowest conduction and the two highest valence bands may be written as
\begin{eqnarray}
\left|\phi_{\text{c}}\right>=\left|d_{z^{2}}\right>,\quad\left|\phi_{\text{v}}\right>=\frac{1}{\sqrt{2}}(\left|d_{x^{2}-y^{2}}\right>-i\left|d_{xy}\right>)
\label{eq:basis},
\end{eqnarray}
where the subscript c/v indicates conduction/valence bands.
It is useful to introduce the Pauli matrix $\hat{\bm{\sigma}}$ to distinguish conduction and valence bands with $\hat{\sigma}_{z}$ defined by $\hat{\sigma}_{z}\left|\phi_{\text{c/v}}\right>=\pm\left|\phi_{\text{c/v}}\right>$.

\subsection{With Mirror Symmetry}\label{sec:theorya}

In the presence of the mirror symmetry, the k$\cdot$p perturbation near the K point results in the effective Hamiltonian $H_{0}$~\cite{D. Xiao 2012},
\begin{eqnarray}
H_{0}&=&\frac{\hat{I}+\hat{\sigma}_{z}}{2}\Big(\varepsilon_{\text{c},\uparrow}(\bm{q})\frac{\hat{I}+\hat{s}_{z}}{2}+\varepsilon_{\text{c},\downarrow}(\bm{q})\frac{\hat{I}-\hat{s}_{z}}{2}\Big) \nonumber\\
&&+\frac{\hat{I}-\hat{\sigma}_{z}}{2}\Big(\varepsilon_{\text{v},\uparrow}(\bm{q})\frac{\hat{I}+\hat{s}_{z}}{2}+\varepsilon_{\text{v},\downarrow}(\bm{q})\frac{\hat{I}-\hat{s}_{z}}{2}\Big) \nonumber\\
&&+\alpha(-q_{x}\hat{\sigma}_{x}+q_{y}\hat{\sigma}_{y})
\label{eq:H0},
\end{eqnarray}
where $\hat{\bm{s}}$ is the Pauli matrix for spin, $\bm{q}$ is the Bloch momentum measured with respect to the K point, and $\varepsilon_{\text{c/v},\uparrow/\downarrow}(\bm{q})$ is quadratic energy-momentum dispersion near the K point for the conduction/valence band with spin $\hat{s}_{z}$=$+1$/$-1$.
Thus $H_{0}$ describes the two lowest conduction bands and the two highest valence bands near the K point [Fig.~\ref{fig:1c}].
The last term in Eq. (\ref{eq:H0}) describes the wavefunction hybridization between $\left|\phi_{\text{c}}\right>$ and $\left|\phi_{\text{v}}\right>$ as $\bm{q}$ moves away from the K point~\cite{D. Xiao 2012}. Recalling that the Berry curvature arises from the $\bm{q}$-dependent change of the wavefunction, the hybridization is crucial for the Berry curvature.
In principle, the Berry curvature arises not only by the hybridization within the conduction and valence bands depicted in Fig.~\ref{fig:1c} but also by the hybridization between $\left|\phi_{\text{c/v}}\right>$ and higher conduction bands and lower valence bands not shown in Fig.~\ref{fig:1c}.
\begin{figure}[ht] 
	\centering 
	\includegraphics[angle=0, width=8.5cm, height=7cm]{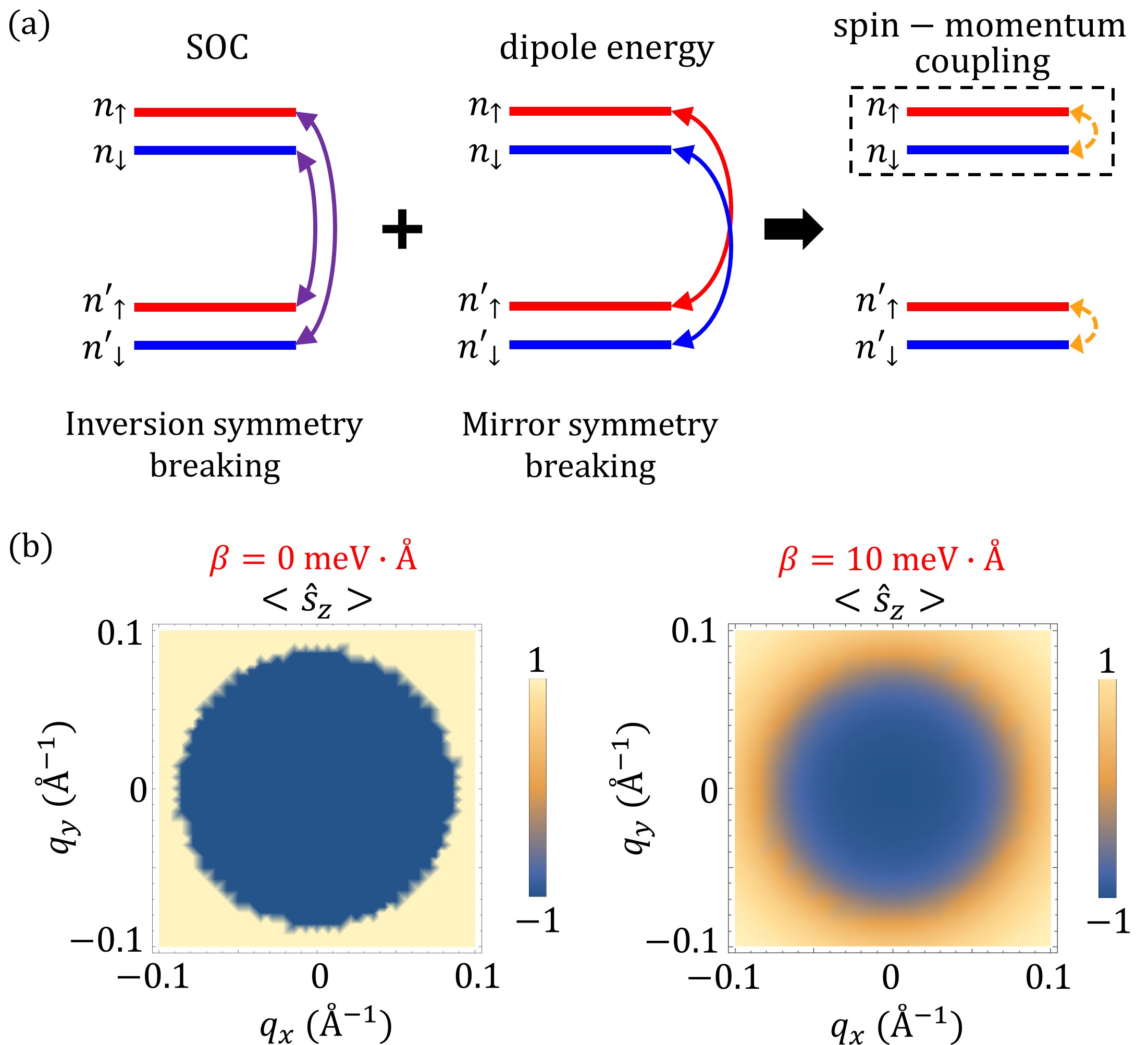} 
	\subfigure{\label{fig:2a}}
	\subfigure{\label{fig:2b}}
\caption{\label{fig:2}
		(Color online)
		(a) Spin-momentum coupling turns up when the mirror symmetry is broken.
		(b) The spin expectation value of the lower conduction band (left) without the mirror symmetry breaking and (right) with the mirror symmetry breaking.
	} 
\end{figure}
However the DFT calculation~\cite{W. Feng 2012} indicates that when the mirror symmetry is present, such hybridization with outer bands has negligible effect on the Berry curvatures of the two lowest conduction bands and the two highest valence bands, although it affects $\varepsilon_{\text{c/v},\uparrow/\downarrow}(\bm{q})$. Thus for the coefficient $\alpha$ in the last term of $H_{0}$, we take its value $\alpha=3.512\,\text{eV}\cdot{\rm \AA}$ from the previous work~\cite{D. Xiao 2012} that captures the hybridization between $\left|\phi_{\text{c}}\right>$ and $\left|\phi_{\text{v}}\right>$ only.
On the other hand, the quadratic dispersion $\varepsilon_{\text{c/v},\uparrow/\downarrow}(\bm{q})$ is chosen in such a way that the energy eigenvalues $E_{\text{c/v},\uparrow/\downarrow}(\bm{q})$ of $H_{0}$ agrees with the band structure from the DFT calculation~\cite{A. Kormanyos 2015}.


\subsection{Without Mirror Symmetry}\label{sec:theoryb}

Until now, we have considered the ideal monolayer MoS$_{2}$ which has the mirror symmetry with respect to its 2D plane. However, the mirror symmetry is broken when the monolayer MoS$_{2}$ is placed on a substrate, which may generate an atomic scale potential gradient (or electric field) and modify the effective onsite and hopping energies of the S atoms in the bottom sublayer of MoS$_{2}$.
These effects can lead to the coupling between the spin $\hat{\bm{s}}$ and the Bloch momentum $\bm{q}$ (spin-momentum coupling)~\cite{Y. A. Bychkov 1984,C. L. Kane 2005,A. Kormanyos 2014,V. Sunko 2017,S. Oh 2017,D. Marchenko 2012,J. Hong 2017}.
Figure~\ref{fig:2a} illustrates the microscopic process by which the spin-momentum coupling may emerge.
$n_{\uparrow/\downarrow}$ denotes one of the four conduction or valence bands shown in Fig.~\ref{fig:1c} whereas $n'_{\uparrow/\downarrow}$ denotes outer bands not shown in Fig.~\ref{fig:1c}, both with the given spin.
The left panel in Fig.~\ref{fig:2a} illustrates the effect of atomic SOC, which induces the spin-orbit interaction between $n_{\uparrow}$ and $n'_{\downarrow}$ and between $n_{\downarrow}$ and $n'_{\uparrow}$.
The middle panel illustrates the effect of the mirror-symmetry-breaking, which induces the hybridization between $n_{\uparrow}$ and $n'_{\uparrow}$, and between $n_{\downarrow}$ and $n'_{\downarrow}$.
The right panel summarizes the combined effect of the atomic SOC and the mirror symmetry breaking; $n_{\uparrow}$ and $n_{\downarrow}$ now couple to each other through the virtual transitions to $n'_{\uparrow}$ and $n'_{\downarrow}$.

As a result, the effective Hamiltonian of the monolayer MoS$_{2}$ with the broken mirror symmetry becomes $H=H_{0}+H_{1}$~\cite{A. Kormanyos 2014}, where
\begin{eqnarray}
H_{1}&=&\frac{\hat{I}+\hat{\sigma}_{z}}{2}\big[\beta_{\text{i}}(\bm{q}\times\hat{\bm{s}})\cdot\hat{\bm{z}}+\beta_{\text{r}}\bm{q}\cdot\hat{\bm{s}}\big]
\label{eq:H1}.
\end{eqnarray}
Here $\beta_{\text{i}}$ and $\beta_{\text{r}}$ are the spin-momentum coupling constants which depend on the degree of mirror symmetry breaking.
Note that in addition to the conventional Rashba spin-momentum coupling $\bm{q}\times\hat{\bm{s}}\cdot\hat{\bm{z}}$, the Weyl spin-momentum coupling $\bm{q}\cdot\hat{\bm{s}}$ coexists. The corresponding couplings for the valence bands are ignored for the reason specified below.

Both the Rashba and Weyl couplings induce $\bm{q}$-dependent spin character change of the wavefunction. Thus the couplings can affect the Berry curvature. For the valence bands, however, this effect is strongly suppressed (and thus ignored) since the $intrinsic$ spin-dependent splitting of the valence bands, $\Delta_{\text{v}}\approx148\,\text{meV}$ [Fig.~\ref{fig:1c}], which exists even without the mirror symmetry breaking, is much stronger than the Rashba and Weyl couplings. For the conduction bands, on the other hand, this effect can be significant since the intrinsic spin-dependent splitting of the conduction bands, $\Delta_{\text{c}}\approx3\,\text{meV}$, is much weaker.


The next section shows that the Berry curvature depends on $\beta_{\rm i}$ and $\beta_{\rm r}$ only through the combination $\beta\equiv\sqrt{\beta_{\rm i}^2+\beta_{\rm r}^2}$. Here we thus estimate $\beta$.
A recent experiment on electron-doped MoS$_{2}$ placed on a SiO$_{2}$ substrate~\cite{J. M. Lu 2015} reports that the product $\beta_{\rm i}k_{\rm F}$ ranges $1\sim 3$ meV (see Fig.~3 in Ref.~\cite{J. M. Lu 2015}), where $k_{\rm F}$ is the Fermi wavelength. Considering that the electron density $n_{\rm 2D}=k_{\rm F}^2/2\pi$ in the experiment is of the order of $10^{14}$ cm$^{-2}$, we conclude that $\beta_{\rm i}$ is of the order of 10 meV$\cdot {\rm \AA}$. For $\beta_{\rm r}$, we do not have any direct estimation but we expect $\beta_{\rm r}$ to be comparable to or smaller than $\beta_{\rm i}$. This leads to the estimation of $\beta\sim 10$ meV$\cdot{\rm \AA}$.
Here we remark that this estimation is at odds with the freestanding monolayer model.
A recent DFT calculation~\cite{A. Kormanyos 2014} examines the effect of a perpendicular electric field $E_{z}$ on a suspended ideal monolayer MoS$_{2}$ and finds $\beta=0.033E_{z}\,\text{eV}\cdot{\rm \AA}$, where $E_{z}$ is in units of $\text{V/\AA}$. Combined with $E_{z}\sim0.03\,\text{V/\AA}$~\cite{Comment} estimated from gate voltages in experiments~\cite{K. F. Mak 2014,J. Lee 2017}, this calculation leads to $\beta\sim 1\ \text{meV}\cdot{\rm \AA}$, which is one order of magnitude smaller than the above estimation obtained from the experiment~\cite{J. M. Lu 2015}. We attribute this difference to the neglect of interatomic hopping between MoS$_{2}$ and substrates (SiO$_{2}$) in the freestanding monolayer model~\cite{A. Kormanyos 2014}.
Recent studies~\cite{J. Hong 2017,S. Oh 2017,V. Sunko 2017} report that interatomic hopping with environment atoms can enhance the spin-momentum coupling strength more than one order of magnitude than estimated from the electric field strength.
As a reference for estimation of this hopping effect, we use results on a monolayer MoS$_{2}$--monolayer graphene heterostructure~\cite{W. Yan 2016,A. Dankert 2017,B. Yang 2017,M. Offidani 2017}, for which it is reported that the graphene acquires the spin-momentum coupling strength of $\sim160\,\text{meV}\cdot{\rm \AA}$~\cite{B. Yang 2017}.
Efficient electrical gate control of spin current is also demonstrated~\cite{A. Dankert 2017,W. Yan 2016}. To obtain the estimation of $\beta$ for our problem, one should take into account the fact that the hopping effect is inverse quadratically proportional to the energy spacing (Fig.~\ref{fig:2}) between bands connected by the hopping, and the energy spacing ($|1.8-8.9|/2=3.6\,\text{eV}$) between MoS$_{2}$ (energy gap $1.8\,\text{eV}$) and SiO$_{2}$ ($8.9\,\text{eV}$) bands is about factor 4 larger than that ($|1.8-0|/2=0.9\,\text{eV}$) between MoS$_{2}$ and graphene ($0\,\text{eV}$) bands. This consideration implies that $\beta$ for the MoS$2$-SiO$_2$ system is about factor $16$ smaller than the corresponding value $\sim 160$ meV$\cdot{\rm \AA}$ for the MoS$_2$-graphene structure. One thus obtains $\beta\sim10\,\text{meV}\cdot{\rm \AA}$ for the MoS$_2$-SiO$_2$ system, which agrees with the above estimation obtained from the experiment~\cite{J. M. Lu 2015}.

We remark that the estimated value of $\beta\sim 10$ meV$\cdot{\rm \AA}$ is about two orders of magnitude smaller than the corresponding value of $\sim 1$ eV$\cdot {\rm \AA}$ in strong spin-momentum coupling systems such as Bi/Ag(111)~\cite{C. R. Ast 2007}. Thus it is reasonable to expect that $H_1$ in Eq.~(\ref{eq:H1}) would generate only weak effects. One example is the spin-conservation violation. Whereas $H_0$ in Eq.~(\ref{eq:H0}) conserves $\hat{s}_z$, $H_1$ does not. But since $\beta$ is very small, the spin conservation is violated only weakly. Figure~\ref{fig:2b} shows that for most values of $\bm{q}$, the expectation value $\langle \hat{s}_z \rangle$ of the lowest conduction bands stay close to $+1$ or $-1$, confirming the weakness of the spin-conservation violation. The spin-conservation induces sizable deviation of $\langle \hat{s}_z \rangle$ from $\pm 1$ only for narrow range of $\bm{q}$ in which the lowest and the second lowest conduction bands become degenerate. However our study in the next section is focused on electronic states very close to the K point ($|\bm{q}|\ll 0.1 {\rm \AA}^{-1}$), so the spin-conservation violation is a weak effect and we use a spin index $s$ to denote eigenstates.


\section{ Result  }\label{sec:result}

In this section, we demonstrate that $H_1$ can induce sizable correction to the Berry curvature even though $\beta$ is small. This becomes possible since the two spin branches of the lowest conduction bands are separated by a small energy spacing of 3 meV. Its demonstration goes as follows.

\begin{figure}[h] 
	\centering 
	\includegraphics[angle=0, width=8.5cm, height=6cm]{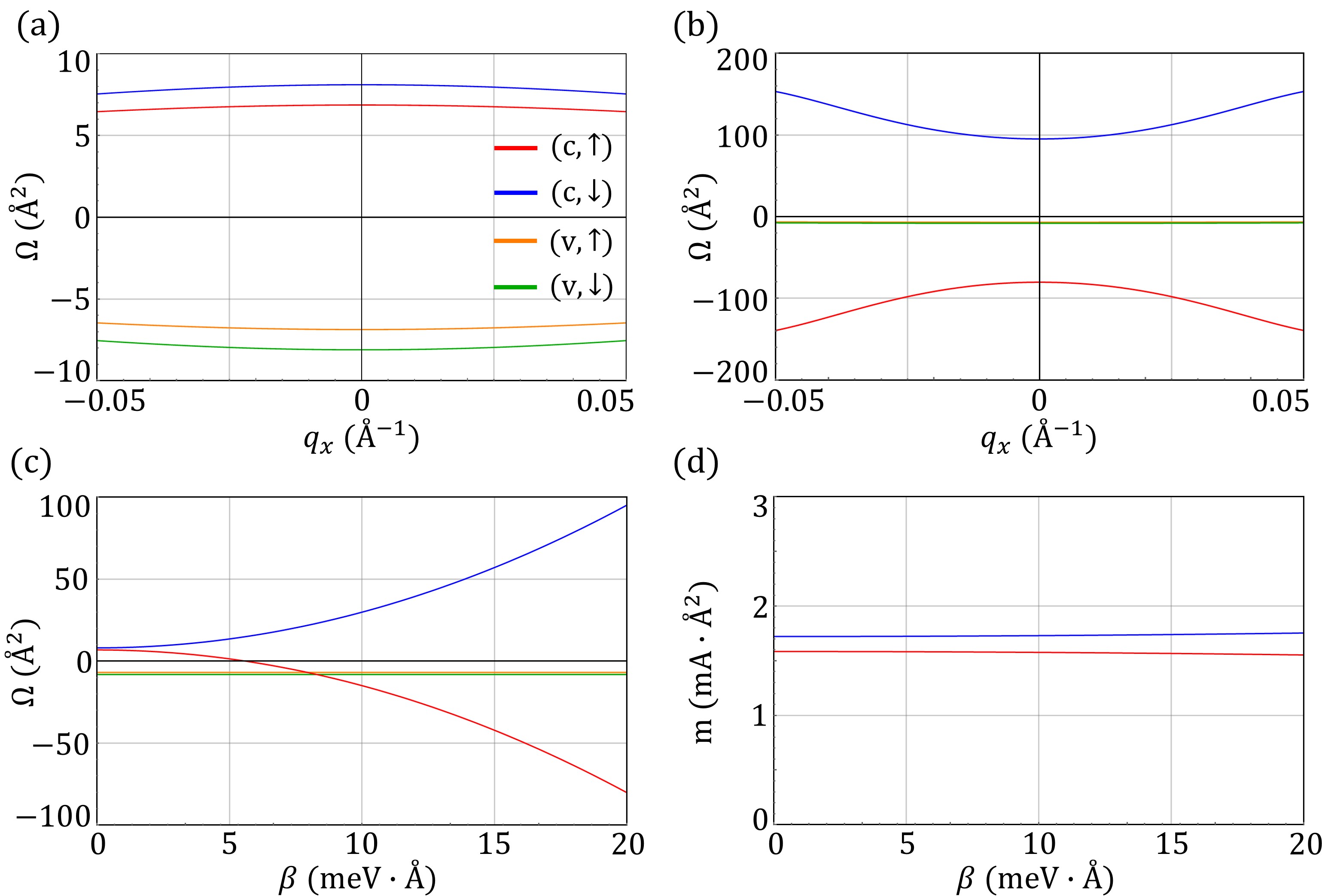} 
	\subfigure{\label{fig:3a}}
	\subfigure{\label{fig:3b}}
	\subfigure{\label{fig:3c}}
	\subfigure{\label{fig:3d}}
	\caption{\label{fig:3}
		(Color online) Calculated values of the Berry curvature near the K point
		(a) with the mirror symmetry ($\beta=0\,\text{meV}\cdot{\rm \AA}$) and (b) without the mirror symmetry ($\beta=20\,\text{meV}\cdot{\rm \AA}$). Note the scale difference in vertical axes of (a) and (b).
		(c) Berry curvature (left) and (d) orbital magnetic moment (right) at the K point as a function of $\beta$.
	} 
\end{figure}
The Berry curvature $\Omega_{n,s}(\bm{q})= \nabla_{\bm{q}}\times i\left<ns\bm{q}\right|\nabla_{\bm{q}}\left|ns\bm{q}\right>\cdot\hat{\bm{z}}$ at each band~\cite{D. J. Thouless 1982,M. C. Chang 1996} is given by
\begin{eqnarray}
\Omega_{n,s}(\bm{q}) &=& i\sum_{\substack{n's'}}'\Bigg[\frac{\left<ns\bm{q}\right|\frac{\partial H}{\partial q_{x}}\left|n's'\bm{q}\right>\left<n's'\bm{q}\right|\frac{\partial H}{\partial q_{y}}\left|ns\bm{q}\right>}{\left[E_{n,s}(\bm{q})-E_{n',s'}(\bm{q})\right]^{2}} \label{eq:Berry}\nonumber\\
&&\qquad\quad-\bigg(\frac{\partial}{\partial q_{x}}\leftrightarrow \frac{\partial}{\partial q_{y}}\bigg)\Bigg]
\label{eq:Berry},
\end{eqnarray}
where $\left|ns\bm{q}\right>$ and  $E_{n,s}(\bm{q})$ are respectively the energy eigenstate and the corresponding energy eigenvalue of $H$, and the summation over $(n',s')$ runs over the four bands of $H$ excluding the case $(n',s')=(n,s)$. The prime symbol ($'$) above $\Sigma$ is introduced to denote this exclusion.
The second line of Eq.~(\ref{eq:Berry}) contains $\partial H/\partial q_{x,y}=\partial H_{0}/\partial q_{x,y}+\partial H_{1}/\partial q_{x,y}$ and allows one to seperate $\Omega_{n,s}(\bm{q})$ into three contributions.
When $\partial H/\partial q_{x}$ and $\partial H/\partial q_{y}$ are replaced by $\partial H_{0}/\partial q_{x}$ and $\partial H_{0}/\partial q_{y}$, one obtains the first contribution, which depends on $\beta_{\text{i}/\text{r}}$ only $implicitly$ through $\left|ns\bm{q}\right>$, $\left|n's'\bm{q}\right>$, $E_{n,s}(\bm{q})$, $E_{n',s'}(\bm{q})$, and captures the hybridization effect between (c, \scriptsize$\uparrow/\downarrow$\normalsize) and (v, \scriptsize$\uparrow/\downarrow$\normalsize) bands (orbital hybridization) due to the last term of $H_{0}$ [Eq.~(\ref{eq:H0})].
When $\partial H/\partial q_{x}$ and $\partial H/\partial q_{y}$ are replaced by $\partial H_{1}/\partial q_{x}$ and $\partial H_{1}/\partial q_{y}$, one obtains the second contribution, which depends $explicitly$ on $\beta_{\text{i}/\text{r}}$ and captures the hybridization effect between (c, \scriptsize$\uparrow$\normalsize) and (c, \scriptsize$\downarrow$\normalsize) bands (spin hybridization) [see Eq.~(\ref{eq:H1})].
``Mixed" contributions coming from $\partial H_{0}/\partial q_{x}$ and $\partial H_{1}/\partial q_{y}$, for instance, vanish since $\partial H_{0}/\partial q_{x}$ and $\partial H_{1}/\partial q_{y}$ induce completely different types of hybridization.
Thus only two contributions survive.
When $(n,s)$ denotes (c, \scriptsize$\downarrow$\normalsize), for instance, straightforward calculation produces
\begin{eqnarray}
\Omega_{\text{c},\downarrow}(\bm{q})&=&\Omega^{\text{cv}}_{\text{c},\downarrow}(\bm{q})+\Omega^{\text{cc}}_{\text{c},\downarrow}(\bm{q}),\label{Be:1}\\
\Omega^{\text{cv}}_{\text{c},\downarrow}(\bm{q})&=&\frac{2\alpha^{2}\Delta}{\big[E_{\text{c},\downarrow}(\bm{q})-E_{\text{v},\downarrow}(\bm{q})\big]^{3}},\label{Be:2}\\
\Omega^{\text{cc}}_{\text{c},\downarrow}(\bm{q})&=&\frac{2\beta^{2}\Delta_{\text{c}}}{\big[E_{\text{c},\downarrow}(\bm{q})-E_{\text{c},\uparrow}(\bm{q})\big]^{3}},\label{Be:3}
\end{eqnarray}
where $\Omega^{\text{cv}}_{n,s}(\bm{q})$ and $\Omega^{\text{cc}}_{n,s}(\bm{q})$ are the Berry curvatures from the orbital and the spin hybridizations, respectively.
$\Omega_{\text{c},\uparrow}(\bm{q})$, $\Omega_{\text{v},\downarrow}(\bm{q})$, and $\Omega_{\text{v},\uparrow}(\bm{q})$ can also be evaluated in a similar way.

The Berry curvatures $\Omega_{\text{c/v},\uparrow/\downarrow}(\bm{q})$ are evaluated for $\beta=0\,\text{meV}\cdot{\rm \AA}$~[Fig. \ref{fig:3a}] and $20\,\text{meV}\cdot{\rm \AA}$ [Fig.~\ref{fig:3b}] as a function of $q_{x}$ with $q_{y}=0$.
Note that $\Omega_{\text{c},\uparrow/\downarrow}(\bm{q})$ is more than one order of magnitude enlarged due to the mirror-symmetry breaking whereas $\Omega_{\text{v},\uparrow/\downarrow}(\bm{q})$ is only very weakly affected. This difference between $\Omega_{\text{c},\uparrow/\downarrow}(\bm{q})$ and $\Omega_{\text{v},\uparrow/\downarrow}(\bm{q})$ stems from $H_{1}$, which affects the wavefunction character only for the conduction bands.
In case of $\Omega_{\text{c},\downarrow}(\bm{q})$, $\Omega^{\text{cc}}_{\text{c},\downarrow}(\bm{q})$ is responsible for the enlargement of $\Omega_{\text{c},\downarrow}(\bm{q})$.
We emphasize that $\Omega^{\text{cc}}_{\text{c},\downarrow}(\bm{q})$ is much larger than $\Omega^{\text{cv}}_{\text{c},\downarrow}(\bm{q})$ not because of its numerator but because of its denominator [Eq.~(\ref{Be:2})];
At the K point, for $\beta=20\,\text{meV}\cdot{\rm \AA}$, its numerator $2\beta^{2}$ is about $(175)^{2}$ times smaller than the numerator $2\alpha^{2}$ of $\Omega^{\text{cv}}_{\text{c},\downarrow}(\bm{q})$, but its denominator $\Delta_{\text{c}}^{2}\sim(3\,\text{meV})^{2}$ is about $(600)^{2}$ times smaller than the corresponding denominator $\Delta^{2}\sim(1.82\,\text{eV})^{2}$.
Their combined effect is more than one order of magnitude enlargement.
Figure~\ref{fig:3c} shows the $\beta$ dependence of the Berry curvature at the K point.
While $\Omega_{\text{v},\uparrow/\downarrow}(\bm{q})$ remains almost independent of $\beta$, $\Omega_{\text{c},\uparrow/\downarrow}(\bm{q})$ grows quadratically as $\beta$ grows.
We also calculate the orbital magnetic moment~\cite{D. Xiao 2007}
\begin{eqnarray}
m_{n,s}(\bm{q}) &=& i\frac{e}{2\hbar}\sum_{\substack{n's'}}'\Bigg[\frac{\left<ns\bm{q}\right|\frac{\partial H}{\partial q_{x}}\left|n's'\bm{q}\right>\left<n's'\bm{q}\right|\frac{\partial H}{\partial q_{y}}\left|ns\bm{q}\right>}{\left[E_{n,s}(\bm{q})-E_{n',s'}(\bm{q})\right]}\nonumber\\
&&\qquad\qquad\,-\bigg(\frac{\partial}{\partial q_{x}}\leftrightarrow \frac{\partial}{\partial q_{y}}\bigg)\Bigg]
 \label{eq:m}.
\end{eqnarray}
 The result is shown in Fig.~\ref{fig:3d}.
Note that the mirror symmetry breaking barely affects the orbital magnetic moment in contrast to its significant effects on the Berry curvature. This difference arises since $m_{n,s}(\bm{q})$ is inversely proportional to the energy difference in contrast to the difference square in case of $\Omega_{n,s}(\bm{q})$. This difference in the energy denominator makes the $\beta$ effect much weaker.

\begin{figure}[ht] 
	\centering 
	\includegraphics[angle=0, width=8.5cm, height=3cm]{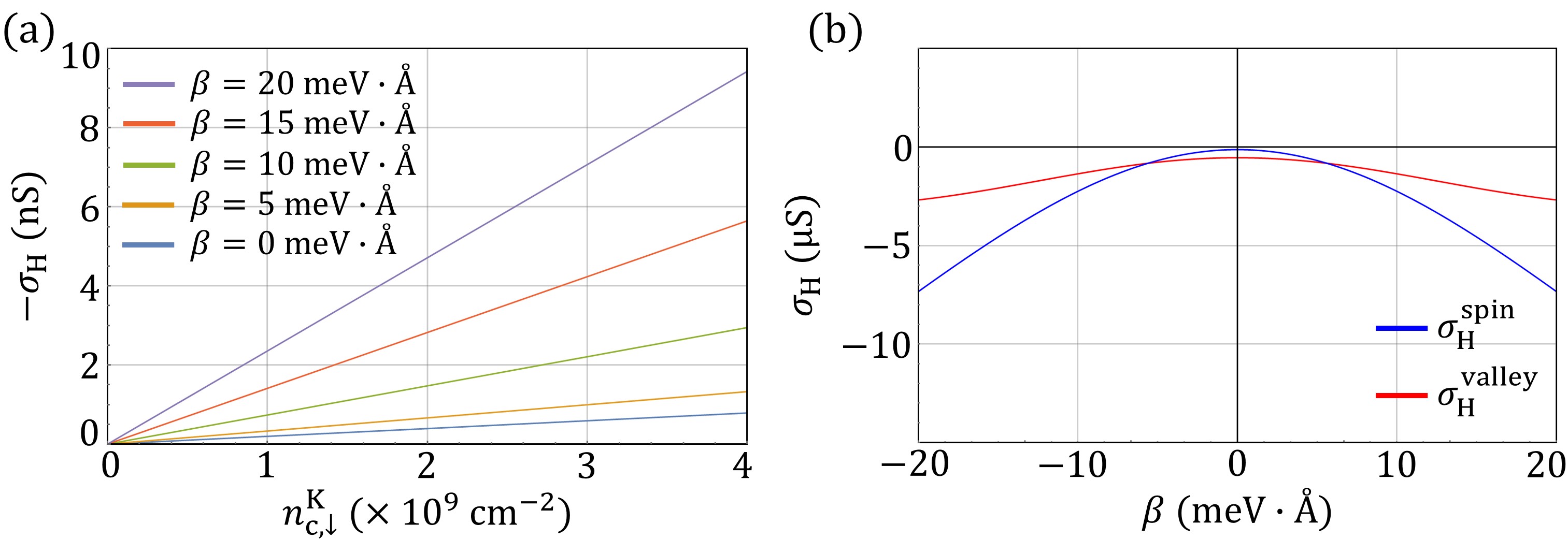} 
	\subfigure{\label{fig:4a}}
	\subfigure{\label{fig:4b}}
	\caption{\label{fig:4}
		(Color online)
		(a) Valley Hall conductivity as a function of the photocarrier density $n^{\text{K}}_{\text{c},\downarrow}$ for various values of $\beta$.
		(b) Spin/Valley Hall conductivity in a monolayer MoS$_{2}$.} 
\end{figure}

Next we compare our calculation results with experiments~\cite{K. F. Mak 2014,J. Lee 2017}.
In the recent experiment~\cite{K. F. Mak 2014}, right/left circularly polarized light is used to selectively excite electrons near the K/K$'$ point from (v, \scriptsize$\uparrow/\downarrow$\normalsize) to (c, \scriptsize$\uparrow/\downarrow$\normalsize) and the Hall conductivity is measured for the optically excited states as a function of the gate voltage and the light intensity, which controls the number of excited carriers.
In the degenerate limit, the Hall conductivity from the intrinsic and side-jump contributions~\cite{D. Xiao 2007,K. F. Mak 2014} is
\begin{eqnarray}
\sigma_{\text{H}}\approx-\frac{e^{2}}{\hbar}\Omega_{\text{c},\downarrow}(0)\cdot n^{\text{K}}_{\text{c},\downarrow}
\label{eq:photo},
\end{eqnarray}
where $n^{\text{K}}_{\text{c},\downarrow}$ is the photocarrier density under the assumption that excitation occurs only to the (c, \scriptsize$\downarrow$\normalsize) band near the K point. The minus sign in Eq.~(\ref{eq:photo}) arises since the side-jump contribution is two times bigger and of opposite sign to the intrinsic contribution. The result is shown in Fig.~\ref{fig:4a}.
Note that (i) $\sigma_{\text{H}}$ is essentially linear in $n^{\text{K}}_{\text{c},\downarrow}$ and (ii) the slope of the $\sigma_{\text{H}}$ vs $n^{\text{K}}_{\text{c},\downarrow}$ curve grows roughly as $\beta^{2}$.
In comparison, the experimental data (Fig.~3 in Ref~\cite{K. F. Mak 2014}) indicates that (iii) $\sigma_{\text{H}}$ grows linearly with the density $\Delta n_{\text{ph}}$ of the photoexcited carriers and (iv) the slope of the linear dependence varies with the gate voltage $V_{\text{g}}$. Here $\Delta n_{\text{ph}}$ and $n^{\text{K}}_{\text{c},\downarrow}$ are related by $n^{\text{K}}_{\text{c},\downarrow}=\Delta n_{\text{ph}}P$, where the ratio $P$ is reported to be much smaller than 1 since it is suppressed by valley relaxation~\cite{H. Z. Lu 2013,Q. Wang 2013,C. Mai 2014,S. D. Conte 2015} and more significantly by the spin relaxation within the conduction bands~\cite{L. Wang 2014,L. Yang 2015}. Such spin relaxation is crucial since $\Omega_{\text{c},\uparrow}(\bm{q})$ and $\Omega_{\text{c},\downarrow}(\bm{q})$ tend to cancel each other [Fig.~\ref{fig:3b}].
If we assume that $P$ remains constant during the experiment, the calculation result (i) agrees with the experimental result (iii). Regarding (ii) vs (iv), we first note that the slope in the experiment is roughly proportional to $(V_{\text{g}}+20)^{2}$. Thus if the effective $\beta$ in the experiment is zero at $V_{\text{g}}=-20\,\text{V}$ and proportional to ($V_{\text{g}}+20\,\text{V}$), the calculation result (ii) also agrees with the experimental result (iv).
As a passing remark, we mention that the “theoretical” curve in Fig.~3 in Ref.~\cite{K. F. Mak 2014} is incorrect by factor 4 due to an error in the Supplementary Material of Ref.~\cite{K. F. Mak 2014}. Once this error is corrected, the curve becomes 4 times steeper and agrees with our curve for $\beta=0\,\text{meV}\cdot{\rm \AA}$ in Fig.~\ref{fig:4a} [except for minor difference due to material parameter difference].

On the other hand, another experiment~\cite{J. Lee 2017}  measured the VHE of an $n$-doped monolayer MoS$_{2}$
in a different way, which does not use the optical excitations but instead utilizes the optical detection (Kerr rotation) of magnetic moment accumulation at edges. The measured dependence on the gate voltage seems to arise mainly from the chemical potental variation and the mirror-symmetry breaking effect appears to be weak.
In the low-energy limit, the corresponding valley Hall conductivity is given by
\begin{flalign}
\sigma_{\text{H}}^{\text{valley}}\!\!=\!
\frac{2e^{2}}{\hbar}\!\!\!\!
\sum_{s=\uparrow/\downarrow}\!\!
\bigg[\!\int\!\!\frac{d\bm{q}}{(2\pi)^{2}}f_{\text{c},s}(\bm{q})\Omega_{\text{c},s}(\bm{q})
\!-\!\frac{q_{\text{\tiny F},s}^{2}}{2\pi}\Omega_{\text{c},s}(q_{\text{\tiny F},s})\bigg]
\label{VHE},
\end{flalign}
where the integration is performed near the K point, the factor 2 is introduced to take into account both the K and K$'$ points, and $f_{\text{c},\uparrow/\downarrow}(\bm{q})$ and $q_{\text{\tiny F},\uparrow/\downarrow}$ are the Fermi distribution function and Fermi momentum for the upper/lower conduction band.
The last term is the side-jump contribution.
Figure~\ref{fig:4b} shows the $\beta$ dependence of $\sigma_{\text{H}}^{\text{valley}}$ in the $n$-doped monolayer MoS$_{2}$. Here we assume that the Fermi energy is fixed (regardless of $\beta$) at $10\,\text{meV}$ above the lower conduction band bottom and the temperature $10\,\text{K}$.
Interestingly the calculated ratio $\sigma^{\rm valley}_{\rm H}/n$, where $n$ is the carrier density in the $n$-doped system, is significantly smaller than the corresponding ratio $\sigma_{\rm H}/n^{\rm K}_{\rm c,\downarrow}$ from Fig.~\ref{fig:4a}; the two ratios are  $-0.32\times 10^{-9}$ nS$\cdot$cm$^{2}$ and $-0.75\times 10^{-9}$ nS$\cdot$cm$^{2}$ for $\beta=10$ meV$\cdot{\rm \AA}$, and $-0.67\times 10^{-9}$ nS$\cdot$cm$^{2}$ and $-2.4\times 10^{-9}$ nS$\cdot$cm$^{2}$ for $\beta=20$ meV$\cdot{\rm \AA}$. This deviation, which gets stronger for larger $\beta$, arises since the both spin-split conduction bands are populated in the optical-detection-based~\cite{J. Lee 2017} scheme and they tend to generate contributions of opposite sign due to the spin hybridization, whereas only one conduction bands are preferrably populated in the optical-excitation-based~\cite{K. F. Mak 2014} scheme and there is no cancellation. Thus the two detection schemes of VHE are not equivalent when the mirror symmetry is broken.


We also calculate the spin Hall conductivity $\sigma_{\rm H}^{\rm spin}$, which can be obtained from Eq.~(\ref{VHE}) by multiplying each term by the proper spin expectation value. Note that for large $\beta$, $\sigma_{\rm H}^{\rm spin}$ is significantly larger than $\sigma_{\rm H}^{\rm valley}$ [Fig.~\ref{fig:4b}] since the two spin-split conduction band contributions now add up.



Another interesting implication of our study is the Rashba-Edelstein effect~\cite{A. Manchon 2015}.
When the mirror symmetry is broken, in-plane chiral spin component arises from the Rashba and Weyl spin-momentum coupling between the two lowest conduction bands in monolayer MoS$_{2}$ [Eq.~(\ref{eq:H1})].
The in-plane spin component and the Berry curvature are maximized at the degenerate points [Fig.~\ref{fig:1c}] with completely hybridized eigenstates.
In such a situation, an electric field can generate spin accumulation (Rashba-Edelstein effect).
A similar enhancement of the Berry curvature has been theoretically proposed \cite{Z. Qiao 2010,W. Tse 2011} for graphene in the context of the quantum anomalous Hall effect.
There is also an experimental paper which reported the change of the inverse Rashba-Edelstein effect depending on the Fermi energy in 2D Rashba system~\cite{E. Lesne 2016}.
A recent experiment~\cite{C. Cheng 2016} on a heterostructure made of a $n$-doped monolayer MoS$_{2}$ and a ferromagnet Co reported large inverse Rashba-Edelstein effect, which may be related to the strong spin-momentum coupling effects near the degenerate points. This relation may be tested experimentally through the material variation since the monolayer MoX$_{2}$ (X=S, Se, Te) all exhibits the degenerate points whereas WX$_{2}$ does not~\cite{A. Kormanyos 2015}.


\section{Discussion}\label {sec:discussion}

Here we discuss a few related issues. The first issue is the substrate effect. In Sec.~\ref{sec:theoryb}, the substrate effect was taken into account through $H_1$. But substrates may generate other types of perturbations as well, which are ignored in this paper. Here we argue that when the coupling with the substrate is weak, the neglect of other perturbations can be a good approximation and $H_1$ describes the most important perturbation in terms of the Berry curvature correction. This point can be seen from the general expression of the Berry curvature in Eq.~(\ref{eq:Berry}). When a perturbation is weak, it usually induces only minor corrections to the Berry curvature. But an exceptional situation can occur when a perturbation induces a hybridization between energy bands with small energy spacing since Eq.~(\ref{eq:Berry}) contains the square of the energy difference in its denominator. In case of MoS$_2$, the two spin branches of the lowest conduction bands are separated by only 3 meV and thus most important perturbations by a substrate are those that hybridize the two spin branches. Considering that the two spin branches have opposite signs of $\hat{s}_z$, perturbations should be able to flip $\hat{s}_z$ to induce the hybridization between the two spin branches. Thus they should contain $\hat{s}_x$ or $\hat{s}_y$. Also considering that $\partial H/\partial q_{x,y}$ appears in the numerator of Eq.~(\ref{eq:Berry}) instead of $H$,  perturbations should depend on $\bm{q}$. Thus near the K (or K') point, most important perturbations are those that are linear in $\bm{q}$ and depend on $\hat{s}_{x}$ or $\hat{s}_{y}$. Note that the two perturbation terms of $H_1$ [Eq.~(\ref{eq:H1})] are exactly of this type, which shows that they are the most important perturbations. This justifies the neglect of other types of perturbations.

The second issue is on the detection of the valley and spin Hall currents. Unlike charge Hall currents, spin Hall currents are not directly observable~\cite{J. Sinova 2006} and as far as we are aware of, there is no direct way to observe valley Hall currents either. One possible option is to infer the valley and spin Hall currents from the accumulated valley and spin densities at side edges of a system. Although this method is adopted in many experiments, the valley and spin Hall conductivities calculated from bulk states [Eq.~(\ref{VHE}) for instance] may not match quantitatively with the valley and spin accumulations at edges since valley and spin are not strictly conserved. Relaxation approximations may be used to quantify the non-conservation effect but this provides only a phenomenological description. In a fundamental level, this issue has not been clearly understood and goes beyond the scope of this paper. We just remark that in case of spin, Ref.~\cite{J. Shi 2006} proposed an alternative definition of a spin current and argued that the modified definition can improve the connection between the bulk spin Hall conductivity and the edge spin accumulation. In Sec.~\ref{sec:result}, we defined a spin current in a conventional way (spin current $\propto$ spin times velocity) since this definition is more commonly used and also the modified definition becomes identical to the conventional definition at the K and K' points ($\bm{q}={\bf 0}$).

Lastly we discuss possible effects of the skew scattering briefly. In case of the anomalous Hall effect, it is well known~\cite{N. Nagaosa 2010} that the skew scattering is the dominant mechanism in very clean systems whereas the intrinsic Berry curvature is dominant in relatively dirty systems. In case of the MoS$_2$ experiment~\cite{K. F. Mak 2014}, the measured Hall conductivity $\sigma_{\rm H}$ is in rough agreement with the prediction of the Berry curvature theory~\cite{D. Xiao 2012} that neglects the mirror symmetry breaking effect. Thus we suspect that the intrinsic Berry curvature is more important in this experiment. This has also motivated us to investigate the deviation between the measured $\sigma_{\rm H}$ and the theory~\cite{D. Xiao 2012} in terms of the Berry curvature modification by the mirror symmetry breaking. However there is a possibility that the skew scattering contributes to the deviation as inferred in Ref.~\cite{K. F. Mak 2014}, although the skew-scattering-based theory of the deviation has not developed yet. Refined experiments are needed to determine whether the mirror symmetry breaking or the skew scattering is the main reason of the deviation.


\section{Summary}\label{sec:summary}

In summary, we calculated the Berry curvature in a monolayer MoS$_{2}$ in realistic situations where the mirror symmetry is broken.
Our focus was not the well-known Rashba spin momentum coupling at the $\Gamma$ point, but the spin momentum coupling at the K point which is a direct band gap point.
We found that the symmetry-breaking contribution to the Berry curvature, which varies with the gate voltage, may be larger than the previously known Berry curvature in an ideal monolayer MoS$_2$ with the symmetry.
However we estimated that the symmetry breaking barely affects the orbital magnetic moment.
This provides an explanation to the gate voltage dependence of VHE in the recent experiment on VHE~\cite{K. F. Mak 2014}.
It also provides an explanation as to why the two recent experiments~\cite{K. F. Mak 2014,J. Lee 2017} show different results with regards to the gate voltage dependence of the VHE.
Large inverse Rashba-Edelstein effect~\cite{C. Cheng 2016}, which is reported recently in a monolayer MoS$_{2}$ and a ferromagnet heterostructure, may be related to our result.

We acknowledge helpful discussion with Jieun Lee and Jonghwan Kim. 
Recently we were informed that Ref.~\cite{B. T. Zhou 2017} reports similar results as ours although it does not discuss its connection with the experiment~\cite{K. F. Mak 2014}. 
Through Ref.~\cite{B. T. Zhou 2017}, we became aware of a recent experiment~\cite{J. M. Lu 2015}, which provides valuable data for the estimation of $\beta$. 
This work was supported by the National Research Foundation of Korea grant (No. 2011-0030046).

\end{document}